# Giant unidirectional magnetoresistance in topological insulator – ferromagnetic semiconductor heterostructures

Nguyen Huynh Duy Khang[1,2] and Pham Nam Hai[1,3,4]

[1]Department of Electrical and Electronic Engineering, Tokyo Institute of Technology,

2-12-1 Ookayama, Meguro, Tokyo 152-0033, Japan

[2]Department of Physics, Ho Chi Minh City University of Education, 280 An Duong Vuong Street,

District 5, Ho Chi Minh City 738242, Vietnam

[3]Center for Spintronics Research Network (CSRN), The University of Tokyo,

7-3-1 Hongo, Bunkyo, Tokyo 113-8656, Japan

[4]CREST, Japan Science and Technology Agency,

4-1-8 Honcho, Kawaguchi, Saitama 332-0012, Japan

**Abstract:** The unidirectional magnetoresistance (UMR) is one of the most complex spin-dependent transport phenomena in ferromagnet/non-magnet bilayers, which involves spin injection and accumulation due to the spin Hall effect (SHE) or Rashba-Edelstein effect (REE), spin-dependent scattering, and magnon scattering at the interface or in the bulk of the ferromagnet. While UMR in metallic bilayers has been studied extensively in very recent years, its magnitude is as small as $10^{-5}$, which is too small for practical applications. Here, we demonstrate a giant UMR effect in a heterostructure of BiSb topological insulator – GaMnAs ferromagnetic semiconductor. We obtained a large UMR ratio of 1.1%, and found that this giant UMR is governed not by the giant magnetoresistance (GMR)-like spin-dependent scattering, but by magnon emission/absorption and strong spin-disorder scattering in the GaMnAs layer. Our results provide new insight into the complex physics of UMR, as well as a strategy for enhancing its magnitude for device applications.

**Introduction**

The UMR effect in ferromagnet (FM)/non-magnet heterostructures with large spin-orbit interaction (SOI) is the only spin-dependent transport phenomenon that breaks the Onsager's reciprocity relations. [1,2,3] Besides new exciting physics, UMR is very attractive for determination of the magnetization direction of a single magnetic layer, making it possible to design novel magnetoresistive random access memories (MRAM) with an extremely simple stacking structure and small foot print. Figure 1(a) shows the schematic structure of a two-terminal planar MRAM device consisting of only a ferromagnetic layer and a spin Hall layer. For data writing, a bipolar pulse current is applied in plane for magnetization switching by the spin-orbit torque that is induced by SHE and/or REE in the spin Hall layer. For data reading, a small bias current can be applied, and a sensing amplifier is used to determine the magnetization direction of the ferromagnetic layer by mean of the UMR effect. Such planar two-terminal memory devices are much simpler than the vertical magnetic tunnel junctions (MTJs) used in conventional MRAM, which involves 30 ultrathin layers including a synthetic antiferromagnetic superlattice and a CoFeB layer for the reference layer, a MgO tunnel barrier, a CoFeB/Ta/CoFeB tri-layer for the free layer, a second MgO layer, and other supporting layers. Furthermore, there is a great degree of freedom in material choice for UMR-MRAM, since there is no restriction to the de facto CoFeB/MgO/CoFeB combination as in the case of MTJs. However, for the UMR effect to be used in such memory devices, its magnitude has to be at least a few % (earliest anisotropy magnetoresistance (AMR)-based MRAMs [4] have a typical AMR ratio of 2% for the Co permalloy as the magnetic layer). Previous studies of UMR in metallic bilayers, such as Ta or Pt/Co,[1,2] reports UMR ratios of about 0.002%, which are too small for practical applications. Recent interests have moved to topological insulator (TI)-based heterostructures, such as $Cr_x(Bi,Sb)_{2-x}Te_3/(Bi,Sb)_2Te_3$,[5]

(Bi,Sb)$_2$Te$_3$/CoFeB and Bi$_2$Se$_3$/CoFeB.[6] Because of the large spin Hall angle in TIs,[7,8,9,10,11] spin accumulation at the interface of those heterostructures can be enhanced, resulting in larger UMR ratios. The physical origins of UMR are explained as combination of spin-dependent scattering (SS) and magnon scattering (MS) mechanisms at the interface or in the bulk of the ferromagnetic layer.[12,13] Figure 1(b) illustrates the SS mechanism in a TI/FM heterostructure. The SHE or REE in the TI layer generates a pure spin current that is injected to the FM layer, resulting in spin accumulation near the surface. The transmission and reflection of spin at the interface depend on direction of the spin polarization $\sigma$ and magnetization $M$, generating a GMR-like UMR signal. Furthermore, since the spin accumulation penetrates to bulk of the FM layer, the bulk spin-dependent scattering (denoted as bulk SS mechanism) of the FM layer also gives rise to a bulk UMR.[14] Note that this bulk UMR can have opposite polarity to the interfacial GMR-like UMR, depending on the mobility of majority/minority-spin electrons. Meanwhile, the MS mechanism involves magnon absorption or stimulated magnon emission in the FM layer when an accumulated spin flips its orientation (Fig. 1(c)). The change of the magnon temperature as the result of the MS process leads to the change of longitudinal resistance via the spin-disorder scattering of free carriers in the FM layer. All of these different mechanisms have recently been observed experimentally,[12] but their magnitude varies with the material combination, and it is not clear how to optimize them to obtain a large enough UMR ratio for device applications. Recently, a large UMR ratio of 0.5% in Cr$_x$(Bi,Sb)$_{2-x}$Te$_3$/(Bi,Sb)$_2$Te$_3$ heterostructures was observed and explained as the manifestation of the enhanced asymmetric magnon scattering that involves the topological surface states with spin-momentum locking in the ferromagnetic TI Cr$_x$(Bi,Sb)$_{2-x}$Te$_3$ layer.[5] However, such asymmetric magnon scattering mechanism is limited to ferromagnetic TIs, whose Curie temperature is much lower than room temperature. Meanwhile, the GMR-like SS is the main mechanism in the (Bi,Sb)$_2$Te$_3$/CoFeB and Bi$_2$Se$_3$/CoFeB heterostructures, [6] but their UMR ratios are much smaller at the order of 10$^{-6}$. Thus, a clear strategy for reaching over 1% of UMR ratio and a comprehensive understanding of the underlying microscopic mechanism of UMR in TI-based heterostructures are strongly required for device applications.

In this work, we investigate the UMR effect and their underlying physical mechanisms in GaMnAs ferromagnetic semiconductor / BiSb topological insulator heterostructures. Here, GaMnAs is a prototype ferromagnetic semiconductor with strong spin-disorder scattering, [15] and BiSb is a topological insulator with a giant spin Hall effect. [10,11] By varying the crystal quality as well as the conductivity of the BiSb layer, we can change the current distribution flowing in the GaMnAs layer and the BiSb layer, and selectively studied the mechanism of UMR in each case. We found that the MS mechanism is dominant in the GaMnAs layer with a maximum UMR of 1.1%, breaking the 1% barrier for the first time. The bulk SS mechanism in the GaMnAs layer is also observed with the corresponding maximum UMR of –0.009%. In addition, we also observed that other factors, such as the anomalous Nernst effect (ANE) of the GaMnAs layer and the ordinary Nernst effect (ONE) of the BiSb layer can dominate the UMR-like signals when the pure spin current injection from the BiSb to the GaMnAs layer is suppressed. Our results provide a strategy to maximise the UMR ratio for practical device applications.

**Preparation of GaMnAs/BiSb heterostructures**

The 10 nm-thick Ga$_{0.91}$Mn$_{0.09}$As (bottom) / 10 nm-thick Bi$_{0.85}$Sb$_{0.15}$ (top) heterostructures were grown on insulating GaAs(001) substrates by molecular beam epitaxy (MBE). The GaMnAs/BiSb heterostructures were grown on semi-insulating GaAs(001) substrates by using an ultrahigh vacuum MBE system. After removal of the surface oxide layer of the GaAs substrate, a 20 nm-thick GaAs buffer was grown to obtain an atomically smooth surface at 580°C. Then, a 10 nm-thick $Ga_{0.91}Mn_{0.09}As$ layer was grown at 300°C with the rate of 9 nm/min. After that, a $Bi_{0.85}Sb_{0.15}$ layer was grown with the rate of 3.5 nm/min. In order to change the BiSb thin film's crystal quality, we grew the BiSb layer at different substrate temperature: 200°C for the sample A and 150°C for the sample B. Finally, the samples were cooled down to the room temperature for deposition of an As thin cap layer. The growth process was monitored *in situ* by reflection high electron diffraction. 100 μm × 25 μm Hall bars along the GaAs[110] direction of the heterostructures were fabricated by optical lithography and ion milling. Contact electrodes consisting of a 60 nm-thick Au and a 6 nm-thick Cr adhesion layer are evaporated on the Hall bars, reducing the Hall bar effective length to 50 μm. Figure 1(d) shows a photograph of a Hall bar with contact electrodes and the measurement configuration. UMR hysteresis curves were measured by sweeping an in-plane external magnetic field $H_{ext}$ along the GaAs[$\bar{1}$10] direction (the *y* direction) under various current density *J*. To extract the UMR hysteresis and eliminate the large even-function component of the conventional magnetoresistance $R_{xx}(H_{ext})$ of GaMnAs, we measured the odd-function component $\Delta R_{xx} = R_{xy}(J) - R_{xy}(-J)$, and the UMR ratio is given by $\Delta R_{xx}^{extr}/R_{xx}(0)$, where $\Delta R_{xx}^{extr}$ and $R_{xx}(0)$ are the extremum of $\Delta R_{xx}(H_{ext} \geq 0)$ and the Hall bar resistance at zero magnetic field, respectively. In addition, to evaluate the contribution of the ANE and the ONE in the UMR signal, we measured $\Delta R_{xx}$ and the odd-function component $\Delta R_{xy} = R_{xy}(J) - R_{xy}(-J)$ of the planar Hall resistance $R_{xy}$ under a rotating in-plane magnetic field.

In order to change the current distribution in the heterostructures, we fabricated two samples, whose the top BiSb layer has different crystal quality and electrical conductivity while the $Ga_{0.91}Mn_{0.09}As$ layer is kept the same. The first sample (denoted as sample A) has a poly-crystalline BiSb layer with low electrical conductivity ($\sigma \sim 1.5 \times 10^4$ $\Omega^{-1}m^{-1}$) matching to that of GaMnAs.[16] Thus, roughly half of the bias current flows in each layer of the sample A. Meanwhile, the second sample (denoted as sample B) has a (001)-oriented BiSb layer[17] with much higher conductivity ($\sigma \sim 2 \times 10^5$ $\Omega^{-1}$m$^{-1}$) than GaMnAs,[18] so that most of the bias current flows in the BiSb layer, while less than 10% of the bias current flows in the GaMnAs. This allows us to study the mechanism of UMR in each case independently. Figure 2(a) and 2(b) show a cross-sectional transmission electron microscope image of sample A and B, respectively, which confirm the different crystal quality of the BiSb layer between two samples. Figure 2(c) shows the temperature dependence of the resistance $R$ of the Hall bar of sample A. The resistance – temperature characteristics of the sample A represents that of GaMnAs, with d$R$/d$T$ singularity at the Curie temperature ($T_C \sim 85$ K).[19] Below $T_C$, $R$ rapidly decreases as the result of suppressed spin-disorder scattering in GaMnAs. Note the large difference between the maximum and the minimum value of $R$ is as large as 26%. The resistance change of sample A is strongly governed the resistance change of the GaMnAs rather than by the BiSb layer. In contrast, the resistance – temperature characteristics of the Hall bar of sample B, shown in Fig. 2(d), represents that of BiSb, showing a plateau at low temperatures due to the dominant metallic conduction on the topological surface states of BiSb.[18] The clear difference between the $R$-$T$ curves of the two samples is consistent with the different distribution of the bias current in the two samples.

**UMR of sample A and its origin**

Figure 3(a)-3(h) show the UMR hysteresis curves $\Delta R_{xx} - H_{ext}$ for the sample A at the base temperature $T_{base}$ of 4 K and the current density $J = 2 \times 10^5$ A/cm$^2$ – $1.6 \times 10^6$ A/cm$^2$. The UMR is negative and has a small magnitude (maximum -0.009%) at small $J \leq 6 \times 10^5$ A/cm$^2$. However, at higher $J \geq 8 \times 10^5$ A/cm$^2$, the UMR becomes positive and increases drastically, reaching 0.55% at $J = 1.6\times10^6$ A/cm$^2$. The switching of UMR polarity as a function of current density indicates that there are competing UMR mechanisms with different sign and $J$-dependency. Considering that the spin Hall angle of BiSb has the same sign as that of Pt, [10,11] the negative UMR observed at small $J$ means that the longitudinal resistance is *larger* when the spin polarization $\sigma$ is parallel to $M$, and *smaller* when $\sigma$ is antiparallel to $M$. This cannot be explained by the interfacial GMR-like SS and the MS mechanism, which should give an ordinary (positive) UMR effect. The only plausible mechanism of this phenomenon is the bulk SS, which can give rise to an inverse UMR when the mobility of the minority electrons is higher than that of majority electrons. [14] Such an inverse UMR effect was recently observed in Co$_{80}$Cr$_{20}$ / Pt bi-layers, where the Co$_{80}$Cr$_{20}$ is known to have higher mobility for minority electrons and the MS mechanism was suppressed by applying a strong external magmatic field. [12] In our case, the bulk SS mechanism in GaMnAs is dominant at low $J$ and low $T_{\text{base}}$, when the MS mechanism is negligible. At elevating $J$, the sample temperature increases due to Joule heating,[20] and the MS mechanism began to take over, resulting in ordinary UMR. To further see the competing action between the bulk SS mechanism and the MS mechanism, we measured the UMR hysteresis at a higher $T_{\text{base}}$ of 30 K, which are shown in Fig. 4(a)-4(h). One can see that the inverse UMR exists only at $J = 2\times10^5$ A/cm$^2$, and switches to the ordinary UMR at $J \geq 6\times10^5$ A/cm$^2$, agreeing with the stronger MS mechanism at higher temperature. Importantly, the UMR ratio reaches 1.1% at $J = 1.6\times10^6$ A/cm$^2$, which is the largest UMR ratio observed so far. We concluded that the dominant mechanism in GaMnAs is MS with the corresponding UMR ratio at the order of 1%, while the bulk SS mechanism also exists but results in a much smaller UMR ratio at the order of $-10^{-5}$.

Figure 5(a) summarizes the UMR ratio as a function of $J$ at various $T_{\text{base}}$. One can see that the UMR ratio rapidly increases with both $J$ and $T_{\text{base}}$. Let the contribution of the bulk SS, the interfacial GMR-like SS, and the MS mechanism to the UMR ratio as $aJ$ ($a < 0$), $bJ$ ($b > 0$), and $cJ+dJ^3$ ($c,d > 0$), respectively. The total UMR ratio can be fitted by $\alpha J + \beta J^3$, where $\alpha = a + b + c$ represents the linear contribution from the bulk SS minus that of the interfacial GMR-like SS and the MS mechanism, while $\beta = d$ represents the non-linear contribution of the MS mechanism at high $J$ due to Joule heating. The fitting shows reasonably good agreement with the experimental results, except for those at $T_{base}$ = 50 K and $J \geq 1.4 \times 10^6$ A/cm², where Joule heating makes the sample temperature close to the Curie temperature (see the Supplementary material[20] for estimation of temperature increase due to Joule heating). Figure 5(b) shows the values of $\alpha, \beta$ at various $T_{base}$. $\alpha$ is negative and nearly unchanged for a wide range of temperature, indicating that the linear component of UMR is dominated by the bulk SS mechanism. Meanwhile, $\beta$ gradually increases up to 50 K. Since UMR due to the MS mechanism depends on magnon absorption/emission, the corresponding $\Delta R_{xx}$ is given by $(dR/dT)*\Delta T_m$, where $\Delta T_m = \gamma J^3$ is the magnon temperature difference between two magnetization directions at high $J$. Thus, $\beta$ is proportional to $\gamma(dR/dT)/R$, where $\gamma$ is independent of temperature as shown later. Our observed $\beta$-$T$ dependence can be explained by the increasing $(dR/dT)/R$ at elevated temperature near the Curie temperature of GaMnAs, as confirmed numerically from the $R$-$T$ data in Fig. 2(c). Figure 5(c) shows $\Delta T_m = \Delta R_{xx} / (dR/dT)$ as a function of $J$ at $T_{base}$ = 30 K (green triangle), 40 K (blue circle), and 50 K (red square). One can see that the $\Delta T_m$ - $J^3$ relationships is universal and can be well fitted by $\Delta T_m = \gamma J^3$ as mentioned above. All of these data confirm that the observed large UMR ratio at high $J$ originates from the MS mechanism.

    The fact that UMR due to the MS mechanism increases with increasing temperature is quite attractive for device applications, because other well-known magnetoresistance effects, such as GMR [21,22] or tunneling magnetoresistance (TMR) [23,24] always decreases with increasing

temperature. To further see the MS mechanism in action, we show the UMR ratio at the small $J$ of $2\times10^5$ A/cm$^2$ as a function of $T_{\text{base}}$ in Fig. 5(d). At $T_{\text{base}} \leq 30$ K, the UMR ratio is still negative since the bulk SS mechanism is dominant at this small current density and low temperatures. However, the UMR ratio becomes positive at $T_{\text{base}} \geq 40$ K, where the non-linear component of the MS mechanism become strong enough.

There is another mechanism that can results in UMR-like behavior of $\Delta R_{xx}$: the anomalous Nernst effect (ANE) of the GaMnAs layer with $V_{\text{ANE}} \propto \boldsymbol{M}\times\nabla T$, where $\nabla T$ is the temperature gradient in the GaMnAs layer. The corresponding thermal voltage is $V_{xx} \propto M_y \nabla T_z$. However, $\Delta R_{xx}$ due to this thermal effect should be proportional to $J$ (since $\nabla T_z \propto J^2$), thus it cannot be the origin of the observed $J^3$ dependence of UMR at high $J$. Nevertheless, we have quantitatively evaluated contribution from the thermal effect by measuring $\Delta R_{xx}$ and $\Delta R_{xy}$ with an in-plane rotating magnetic field. Figure 6(a) and 6(b) show the magnetic field direction $\theta$-dependence of $\Delta R_{xx}$, measured at $T_{\text{base}} = 4$ K, $J = 1.6\times10^6$ A/cm$^2$, and $H_{\text{ext}} = 2$ kOe and 7.8 kOe, respectively. The amplitude of the $\Delta R_{xx}$-$\theta$ curve (~ 20 Ω) at $H_{\text{ext}} = 7.6$ kOe is smaller than that (~ 40 Ω) at $H_{\text{ext}} = 2$ kOe due to suppression of magnon population under a higher magnetic field, as expected from the dominant MS mechanism. Figure 6(c)-6(e) show the representative $\theta$-dependence of $\Delta R_{xy}$, measured at $H_{\text{ext}} = 2$ kOe, 4 kOe, and 7.8 kOe, respectively. The experimental $\Delta R_{xy}$-$\theta$ curves can be decomposed to the field-like torque's contribution (green curves) $R_{\text{FL}} = R_{\text{PHE}}\times(H_{\text{FL}}/H_{\text{ext}})\times(\sin3\theta - \sin\theta)$, where $R_{\text{PHE}}$ is the planar Hall resistance and $H_{\text{FL}}$ is the field-like effective field, and the antidamping-like torque and thermal effects' contribution (blue curves) $R_{\text{AD}} + R_{\nabla T} = (c_{\text{AD}}/H_{\text{ext}} + c_{\text{AN}} + c_{\text{ON}}\times H_{\text{ext}})\sin\theta$, where $c_{\text{AD}}$, $c_{\text{AN}}$ and $c_{\text{ON}}$ are the coefficient of antidamping-like torque, the ANE of the GaMnAs layer, the ONE of the BiSb layer, respectively, and $H_{\text{AD}}$ is the antidamping-like effective field. By fitting to the experimental $\Delta R_{xy}$-$\theta$ curves, we can extract $R_{AD} + R_{\nabla T}$. Figure 6(f) shows the extracted $R_{AD} + R_{\nabla T}$ amplitude as a function of $H_{ext}$, which yields $c_{AN}$ = 0.46 Ω and $c_{ON}$ = 0.02 Ω·kOe$^{-1}$, respectively. Thus, ANE is the main thermal effect, and its contribution to $\Delta R_{xx}$ is 0.92 Ω at $J = 1.6 \times 10^6$ A/cm$^2$, estimated from the Hall bar's size. The corresponding UMR ratio due to the ANE effect in sample A is about 0.02% at $T_{base}$ = 4 K and $J = 1.6 \times 10^6$ A/cm$^2$, which is much smaller than the observed UMR ratio of 0.56%, and even smaller than the estimated -0.056% from the bulk-SS mechanism. We conclude that while there are thermal effects contribution to UMR, its magnitude is negligible in sample A.

**UMR of sample B and its origin**

In contrast to sample A where the conductivity of the BiSb layer and the GaMnAs layer is nearly the same, the conductivity of the BiSb layer is much higher than that of the GaMnAs layer in sample B. Such a situation is suitable to study the role of the conductivity matching in the UMR effect. Figure 7(a)-7(h) show the UMR hysteresis curves $\Delta R_{xx}$ – $H_{ext}$ of sample B at $T_{base}$ = 4 K and $J = 2 \times 10^5$ A/cm$^2$ – $1.6 \times 10^6$ A/cm$^2$ for sample B. In this sample, most of the bias current flows in the BiSb layer, thus the bulk SS and MS mechanism in the GaMnAs are not detected. In contrast to sample A, the UMR ratio of sample B increases slowly, reaching a maximum UMR ratio of 0.020% at $J = 1.6 \times 10^6$ A/cm$^2$, which is an order of magnitude smaller than that of sample A measured at the same condition. In addition, we observed a linear dependence with a positive gradient of UMR on the external magnetic field $H_{ext}$ at high $J$, indicating contribution of the ONE. Figure 8(a) summarizes the UMR ratio of sample B as a function of $J$ at various $T_{base}$. The UMR ratio increases linearly with $J$ at all temperatures, and its maximum value at $J = 1.6 \times 10^6$ A/cm$^2$ is nearly temperature-independent. At the first look, this behavior seems consistent with the interfacial GMR-like SS mechanism. Here, we show that this UMR-like signal in the sample B is completely governed by the thermal effects. Figure 8(b) shows the $\theta$ - dependence of $\Delta R_{xy}$ for sample B at $J = 1\times 10^5$ A/cm² with $H_{ext}$ = 1.0 and 7.6 kOe. One can see that $\Delta R_{xy}$ is dominated by the $(R_{AD} + R_{\nabla T})\sin\theta$ component that increases with increasing of $H_{ext}$, while the field-like torque $R_{FL}$ component is absent. Figure 8(c) shows the extracted $R_{AD} + R_{\nabla T}$ amplitude as a function of $H_{ext}$ at $J = 1\times 10^5$ and $2\times 10^5$ A/cm², which reveals that the antidamping-like torque $R_{AD}=c_{AD}/(H_{ext} + H_{AD})$ component is also absent, and the thermal component $R_{\nabla T}= c_{AN} + c_{ON}\times H_{ext}$ is dominant. At $J = 2\times 10^5$ A/cm², we obtained $c_{AN}$ = 0.075 Ω and $c_{ON}$ = 0.016 Ω·kOe⁻¹. The corresponding contribution of ANE and ONE to $\Delta R_{xx}$ are 0.15 Ω and 0.032 Ω·kOe⁻¹ at $J = 2\times 10^5$ A/cm². Thus, the thermal effects perfectly match the UMR hysteresis in Fig. 7(a), which means there is no UMR related to the GMR-like SS mechanism. This is consistent with the absence of the pure spin current injection from the BiSb layer to the GaMnAs layer, as evidenced by the zero field-like torque and antidamping-like torque effects in the $\Delta R_{xy}-\theta$ curves. The absence of pure spin current injection in sample B can be explained using the spin accumulation theory. [2] The spin accumulation at the BiSb/GaMnAs interface can be expressed as

$$\mu_s \approx \mu_s^0 \tanh\frac{t_N}{2\lambda_N} \frac{1}{1+(1-P^2)\frac{\rho_N\lambda_N}{\rho_F\lambda_F}\coth\frac{t_N}{\lambda_N}\tanh\frac{t_F}{\lambda_F}} \qquad (1)$$

where $\mu_s^0 = 2e\rho_N\lambda_N\theta_{SH}J$ is the bare spin accumulation due to the SHE that would occur in a single and infinitely thick BiSb layer, $\rho_N$ and $\rho_F$ are the resistivity, $\lambda_N$ and $\lambda_F$ are the spin diffusion length, $t_N$ and $t_F$ are the thickness of the BiSb and GaMnAs layer, respectively, and $P$ is the spin polarization of GaMnAs. Because the spin diffusion length of GaMnAs is two orders of magnitude larger than that of BiSb, [25,26] $\frac{\rho_N \lambda_N}{\rho_F \lambda_F}$ is nearly zero. Thus, Eq. (1) can be simplified as

$\mu_s \approx \mu_s^0(\rho_N)\tanh\frac{t_N}{2\lambda_N}$, which depends on the resistivity of the BiSb layer. Since the resistivity of sample B is an order of magnitude smaller than that of sample A, sample B has much smaller spin accumulation and spin injection at the BiSb/GaMnAs interface than sample A. This problem is similar to the "conductivity mismatch" problem, which is known to prevent spin injection in bilayers with significantly different conductivity, such as metallic ferromagnet/semiconductor heterostructures. [27]

**Discussion**

Our results demonstrate that it is possible to obtain a UMR ratio larger than 1% in a ferromagnetic semiconductor / topological insulator heterostructures, which is larger than those in metallic bilayers and TI / metallic ferromagnet systems by several orders of magnitude. In particular, our results show that utilizing magnon scattering in ferromagnetic materials with strong spin disorder scattering and conductivity matching are the key factors to obtain a large UMR ratio for UMR-MRAM applications. Besides the required strong spin disorder scattering, the ferromagnetic material should have high enough resistivity to reduce the effect of parasitic resistance of electrical interconnection. Ferromagnetic semiconductors, like the GaMnAs material used in this work, are promising. Although the highest Curie temperature of GaMnAs thin films reported so far (~ 200 K) [28] is still lower than room temperature, several Fe-doped narrow-gap ferromagnetic semiconductors with room-temperature ferromagnetism, such as GaFeSb [29] or InFeSb [30,31], have been realized recently. On the other hand, topological insulators, such as $Bi_2Se_3$, [7] $(BiSb)_2Te_3$, [8] sputtered $Bi_xSe_{1-x}$, [9] or poly-crystalline BiSb (this work), are promising as the spin Hall layer, since they have large spin Hall angle and comparable electrical conductivity with ferromagnetic semiconductors. In contrast to GMR or TMR effects for which material combinations have been exhaustedly explored, those for UMR are largely unexplored. Further material engineering may improve the UMR ratio to over 10%, which is essential for UMR-MRAM with extremely simple structure and fast reading. Our demonstration of an UMR ratio over 1% is an important step toward this goal.

**Figures and Captions**

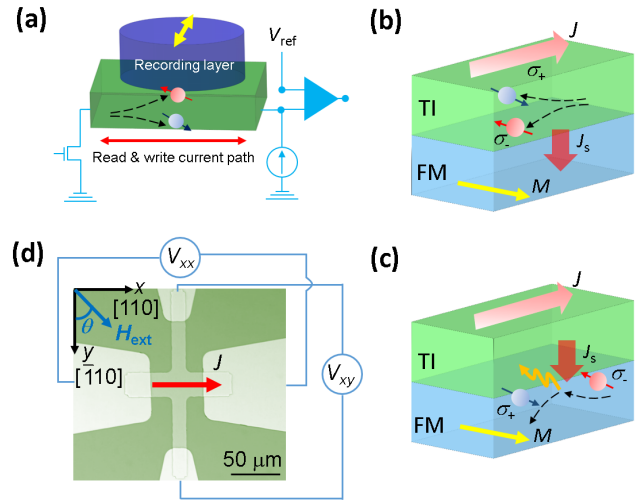

**Figure 1.** (a) Two-terminal planar MRAM device with an extremely simple stacking structure and small footprint, consisting of only a ferromagnetic layer and a spin Hall layer. Spin-orbit torque and unidirectional magnetoresistance are utilized for data writing and reading, respectively. (b), (c) Schematic illustration of the spin-dependent scattering (SS) and magnon scattering (MS) mechanism of UMR, respectively. (d) Optical image of a $Ga_{0.91}Mn_{0.09}As$ (10 nm)/$Bi_{0.9}Sb_{0.1}$ (10 nm) Hall bar and magnetoresistance measurement configuration.

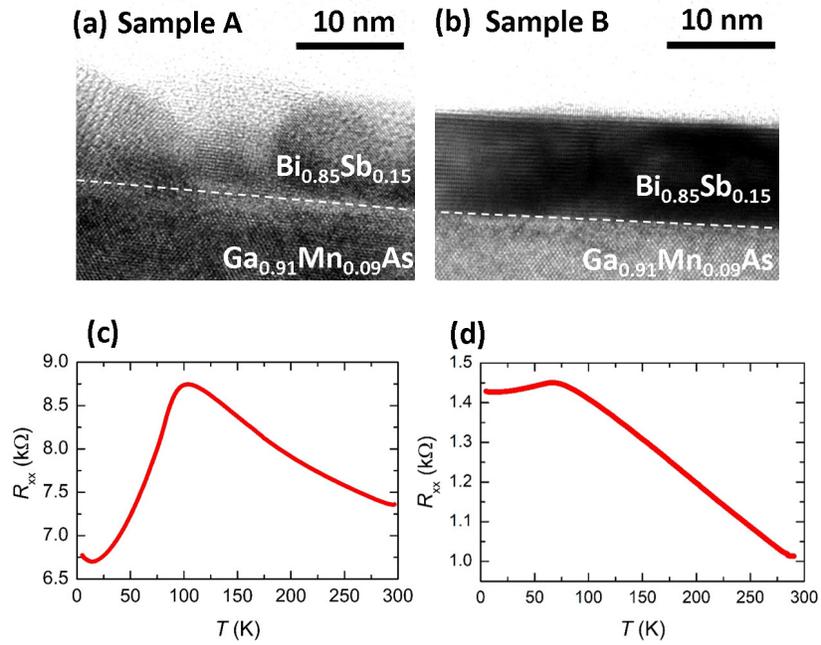

**Figure 2.** (a)(b) Cross-sectional transmission electron microscope image of sample A and B, respectively. The top BiSb layer is polycrystalline with the electrical conductivity of about $1.5\times10^4$ $\Omega^{-1}$m$^{-1}$ in sample A, while it is (001)-oriented with the electrical conductivity of about $2\times10^5$ $\Omega^{-1}$m$^{-1}$ in sample B. (c)(d) Temperature dependence of the resistance of a Hall bar of sample A and B, respectively.

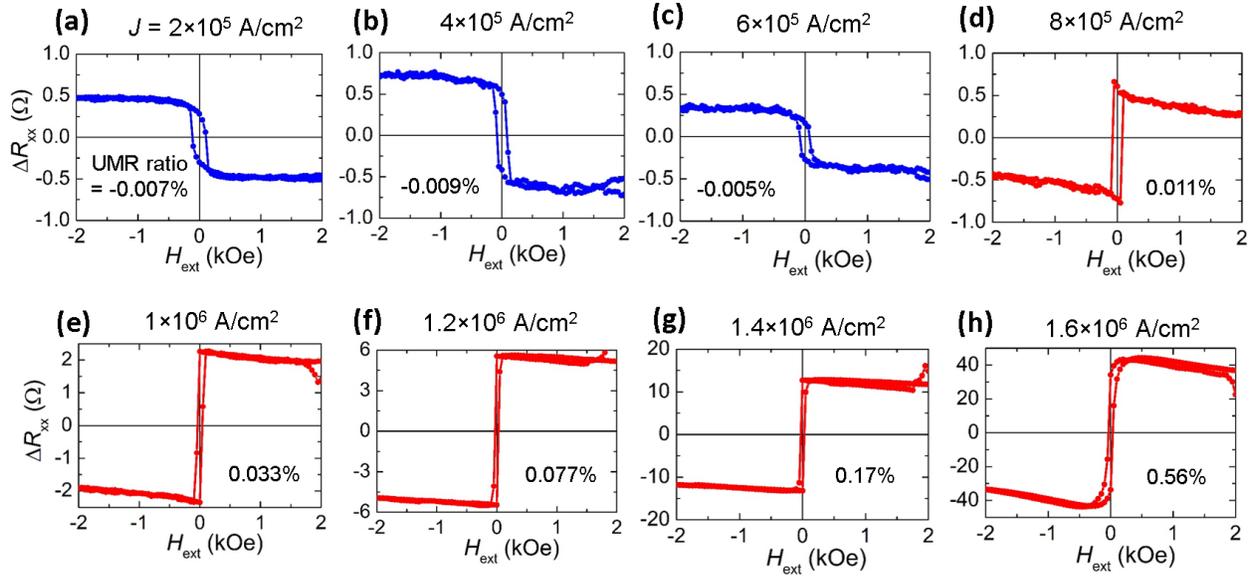

**Figure 3.** UMR hysteresis curves $\Delta R_{xx} - H_{\text{ext}}$ for sample A at the base temperature $T_{\text{base}}$ of 4 K and the current density $J = 2\times 10^5$ A/cm$^2$ – $1.6\times 10^6$ A/cm$^2$. The magnetic field was applied along the $y$ direction. The inverse (negative) UMR at low $J$ is governed by the bulk SS mechanism, while the ordinary (positive) UMR at high $J$ is due to the MS mechanism.

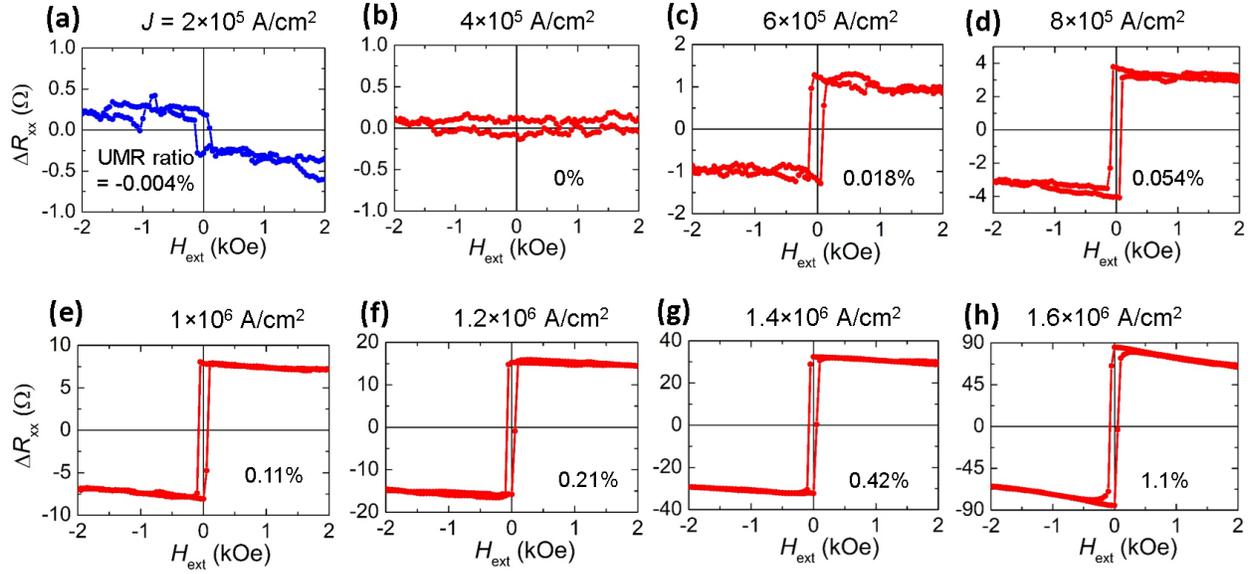

**Figure 4.** UMR hysteresis curves at 30 K for sample A. The inverse UMR exists only at $J = 2\times10^5$ A/cm$^2$, and switches to the ordinary UMR at $J \geq 6\times10^5$ A/cm$^2$ due to stronger MS mechanism at higher temperature. The UMR ratio reaches 1.1% at $J = 1.6\times10^6$ A/cm$^2$, breaking the 1% barrier.

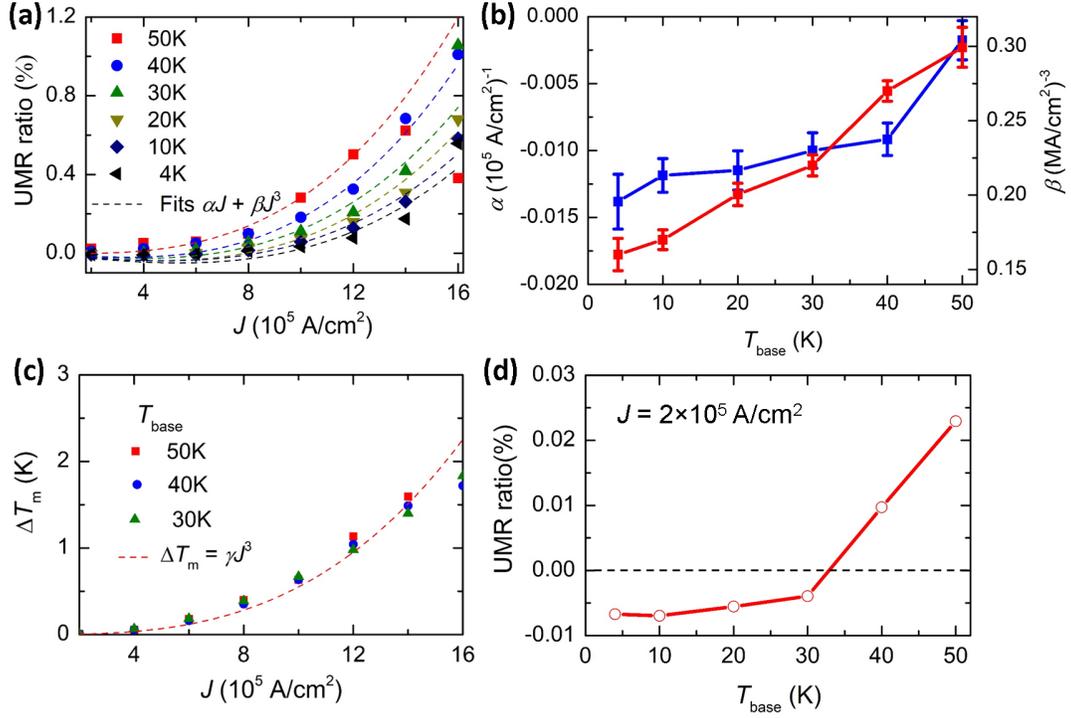

**Figure 5.** (a) UMR ratio as a function of $J$ at various $T_{base}$. Dashed lines show the fitting curves by $\alpha J + \beta J^3$. (b) Values of $\alpha$ (blue) and $\beta$ (red) at various $T_{base}$. $\alpha$ is negative and nearly unchanged for a wide range of temperature, indicating that the linear component of UMR is dominated by the bulk SS mechanism. Meanwhile, $\beta$ gradually increases up to 50 K, consistent with the MS mechanism. (c) Estimation of the magnon temperature difference $\Delta T_m$ as a function of $J$ at difference base temperature. (d) UMR ratio at the very small $J$ of $2\times10^5$ A/cm$^2$ as a function of $T_{base}$ in Fig. 5(c), reflecting the competition between the bulk SS and MS mechanism.

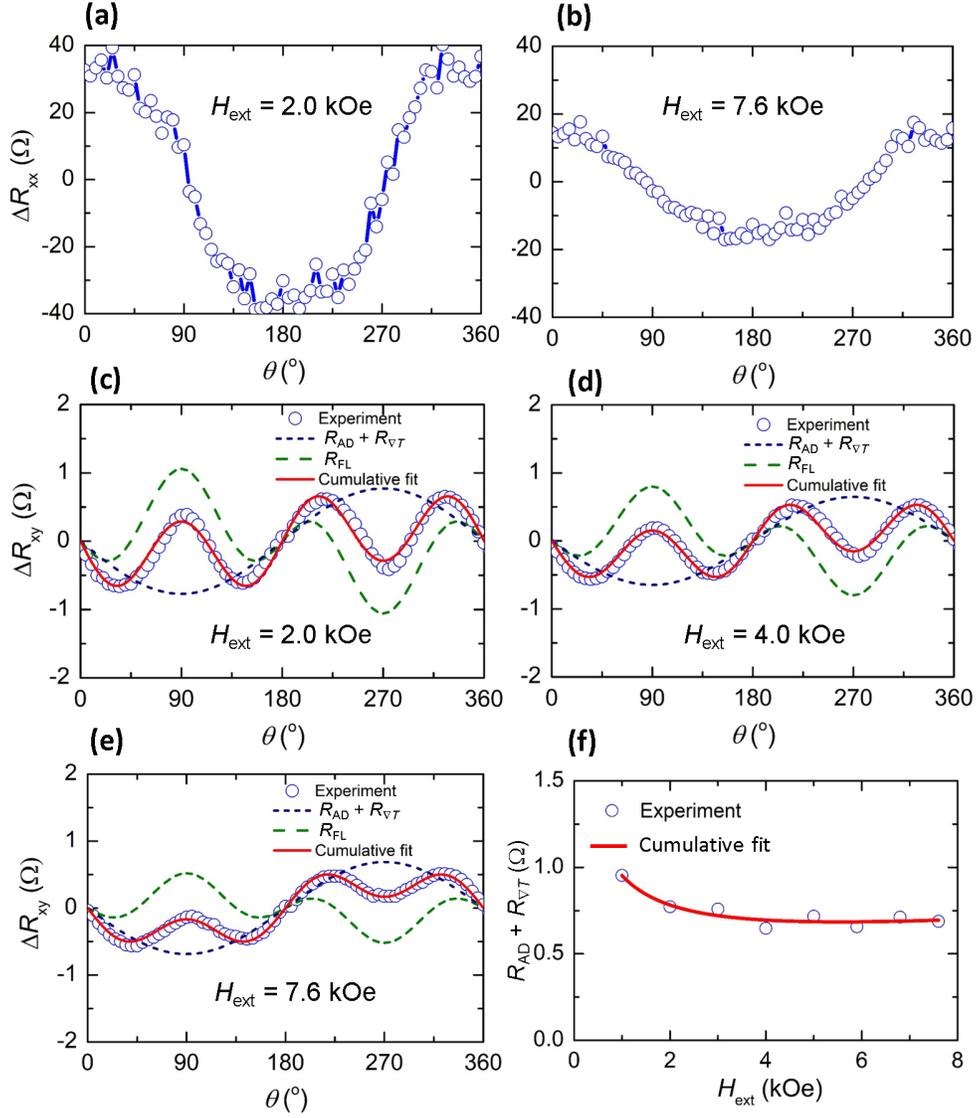

**Figure 6.** Contribution of thermal effects to UMR in sample A. (a), (b) Magnetic field direction $\theta$-dependence of $\Delta R_{xx}$, measured at $T_{base}$ = 4 K, $J$ = 1.6×10$^6$ A/cm$^2$, under $H_{ext}$ = 2.0 and 7.6 kOe, respectively. (c), (d) $\theta$-dependence of $\Delta R_{xy}$, measured at $T_{base}$ = 4 K, $J$ = 1.6×10$^6$ A/cm$^2$, and $H_{ext}$ = 2.0 and 7.6 kOe, respectively. Blue circles: experimental data, green curves: contribution from the field-like torque $R_{FL}$, blue curves: contribution from the antidamping-like torque and thermal effects $R_{AD} + R_{\nabla T}$, red curves: cumulative fitting curves. (e) $R_{AD} + R_{\nabla T}$ amplitude as a function of $H_{ext}$. Red line shows the fitting curve by $c_{AD}/(H_{ext} + H_{AD}) + c_{AN} + c_{ON} \times H_{ext}$.

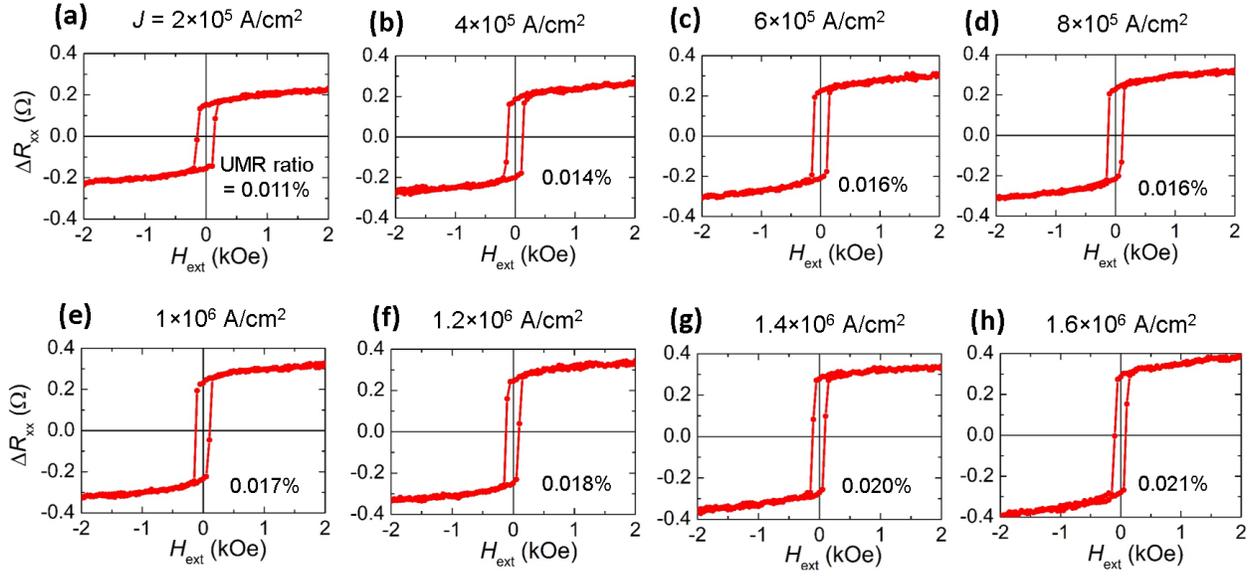

**Figure 7.** UMR hysteresis curves $\Delta R_{xx} - H_{ext}$ of sample B at $T_{base}$ = 4 K and $J = 2\times10^5$ A/cm$^2$ – $1.6\times10^6$ A/cm$^2$ for sample B.

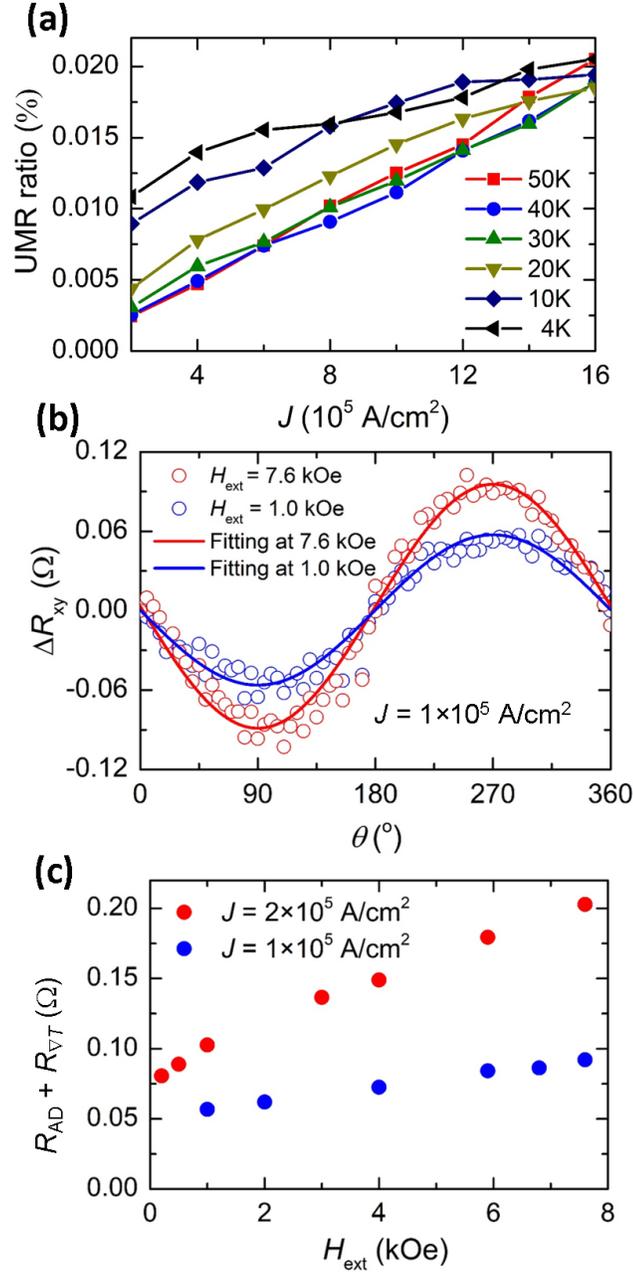

**Figure 8.** (a) UMR ratio of sample B as a function of $J$ at various $T_{base}$. (b) $\theta$−dependence of $\Delta R_{xy}$, measured at $T_{base}$ = 4 K, $J$ = 1×10$^5$ A/cm$^2$, and $H_{ext}$ = 1 kOe (blue) and 7.6 kOe (red), respectively. Solid lines show the fitting curves by the contribution from $R_{AD} + R_{\nabla T}$. (c) $R_{AD} + R_{\nabla T}$ amplitude as a function of $H_{ext}$ at $J$ = 1×10$^5$ A/cm$^2$ (blue) and 2×10$^5$ A/cm$^2$ (red).

Supplemental Material for

# Giant unidirectional magnetoresistance in topological insulator – ferromagnetic semiconductor heterostructures

Nguyen Huynh Duy Khang[1,2] and Pham Nam Hai[1,3,4*]

[1]Department of Electrical and Electronic Engineering, Tokyo Institute of Technology,

2-12-1 Ookayama, Meguro, Tokyo 152-8550, Japan

[2]Department of Physics, Ho Chi Minh City University of Education, 280 An Duong Vuong

Street, District 5, Ho Chi Minh City 738242, Vietnam

[3]Center for Spintronics Research Network (CSRN), The University of Tokyo,

7-3-1 Hongo, Bunkyo, Tokyo 113-8656, Japan

[4]CREST, Japan Science and Technology Agency,

4-1-8 Honcho, Kawaguchi, Saitama 332-0012, Japan

*Corresponding author: pham.n.ab@m.titech.ac.jp

**Estimation of Joule heating in sample A**

Figure S1 shows estimation of temperature increase due to Joule heating in sample A. These temperature was extracted by comparing the temperature dependence of its longitudinal resistance (Fig. 2(c)) and its real resistance at a given $J$ and $T_{base}$. At $J = 16\times10^5$ A/cm$^2$ and $T_{base} = 40$ K, $T \sim$ 83 K, which is close to the Curie temperature $T_C \sim 85$ K. At $J = 16\times10^5$ A/cm$^2$ and $T_{base} = 50$ K, $T \sim 89$ K that leads to decrease of the UMR ratio. In sample B, we didn't observe the Joule heating effect even at $J = 16\times10^5$ A/cm$^2$.

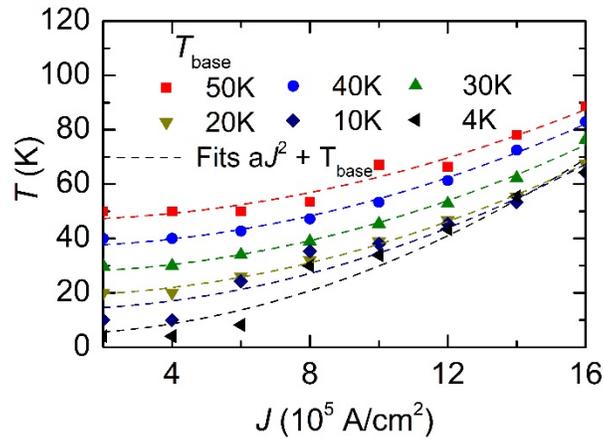

**Figure S1.** Real temperature $T$ of sample B as functions of current density $J$ at different base temperatures $T_{base}$.